\documentclass[aps,prb,twocolumn,nopacs,amsmath,amssymb]{revtex4-1}

\usepackage[dvips]{graphicx}
\usepackage{dcolumn}
\usepackage{bm}
\usepackage{color}
\def\be{\begin{equation}}
\def\en{\end{equation}}
\def\bea{\begin{eqnarray}}
\def\ena{\end{eqnarray}}
\def\bec{\begin{equation}\begin{array}{rcl}}

\def\p{\partial}
\def\ep{\epsilon}
\def\gs{\gtrsim}
\def\ls{\lesssim}

\def\ve{\varepsilon}
\newcommand{\av}[1]{\langle{#1}\rangle}

\newcommand{\bi}[1]{\mbox{\boldmath$#1$}}
\newcommand{\pp}[2]{\frac{\partial {#1}}{\partial {#2}}}

\def\hrij{\hat{\bi r}_{ij}}

\def\aQ{\stackrel{\leftrightarrow}{Q}}
\def\aI{\stackrel{\leftrightarrow}{I}}

\begin{document}
\title{
Structural   phase transition 
and   orientation-strain glass formation 
in anisotropic particle systems with 
impurities in two dimensions 
}  
\author{Kyohei Takae and Akira Onuki}
\affiliation{Department of Physics, Kyoto University, Kyoto 606-8502, Japan}


\date{\today}

\begin{abstract} 
Using  a modified Lennard-Jones 
model for elliptic  particles and spherical  
impurities, we 
present results of molecular dynamics simulation 
in two dimensions.   
In one-component systems of elliptic particles,  
we find    an orientation  phase  transition 
on a hexagonal lattice 
as the temperature $T$ is lowered. 
It is also   
   a structural one because of 
   spontaneous  strain.  
At low  $T$, there arise   three martensitic variants 
 due to the underlying lattice, 
leading to a  shape memory effect without dislocation 
formation. Thermal hysteresis,  
  a minimum of  the shear modulus,  
and a maximum of the specific heat 
are also found with varying $T$.   
With increasing the composition $c$ of 
  impurities, 
the three kinds of orientation 
domains are finely divided, 
 yielding orientation-strain 
 glass with mesoscopically ordered  regions   still surviving. 
If the impurities are large and repulsive, 
planar anchoring of the elliptic particles occurs around 
the impurity surfaces. 
If they are small and attractive, 
homeotropic anchoring occurs.  
Clustering of impurities is conspicuous.  
With increasing the  anchoring  power 
and/or  the composition of the impurities, 
 positional  disorder can  also be enhanced. 
We also investigate the rotational dynamics of 
the molecular orientations. 
\end{abstract}



\maketitle


\section{Introduction}

Certain anisotropic molecules  such as N$_2$, C$_{60}$, 
and KCN form 
a cubic crystal and, at  lower  temperatures, 
they undergo  an orientation  phase transition 
with a specific-heat peak \cite{ori,Yamamuro}, 
where the crystal structure  changes 
 to  a noncubic one. 
Furthermore,  mixtures of anisotropic particles 
\cite{ori} 
such as (KCN)$_x$(KBr)$_{1-x}$     and 
one-component systems of 
globular molecules \cite{Yamamuro} such as ethanol 
and   cyclohexanol   become orientation glass. 
In such glass, 
the phase ordering   should occur only on 
 small spatial scales with  
 mesoscopically  heterogeneous   
orientation fluctuations.
 Because of  anisotropic molecular shapes, 
 there should be a direct (proper) coupling between 
the   molecular orientations and 
the lattice deformations \cite{Bell}. 
In fact, the shear modulus  becomes small 
 around  the orientational   
order-disorder  or glass transition 
 \cite{sound,sound1}. 
These systems thus  exhibit 
singular acoustic and plastic 
 behaviors   \cite{Sherwood}, but 
 there has been no 
systematic  experiment  in the nonlinear response 
regime.  Many of these anisotropic molecules  have dipolar moments also, 
yielding dielectric anomaly near the transition.

In metallic alloys,  a structural phase transition 
arises from  the displacements of 
the atoms  in each unit cell 
 from their equilibrium positions in the 
high-symmetry phase. 
Some alloys undergo a martensitic  
phase transition gradually from a high-temperature phase  
to a low-temperature phase 
over a rather wide temperature range 
and, at sufficiently  low temperatures, 
they  are composed of 
multiple martensitic  variants or domains 
 \cite{marten1,marten2,marten3,RenReview}.  
In particular,   a system of  
 off-stoichiometric intermetallic  Ti$_{50-x}$Ni$_{50+x}$    
has been    studied extensively 
\cite{marten3,RenReview,Ren}.  Even at $x=1.5$, it   
 becomes    strain glass,  exhibiting   the shape-memory effect 
and the superelasticity, 
where strain heterogeneities with  sizes of order 
10 nm were observed \cite{Ren}. 
As a similar example, 
metallic ferroelectric glass, called  relaxor,  
exhibits  large dielectric response 
to applied electric field   
\cite{relax,CowleyReview,Hirota}, 
where  the  electric polarization 
 and the lattice deformations are coupled  
 and frozen polar nanodomains 
are produced in the presence of the compositional disorder 
in the  perovskite structure.

In  soft matter, 
impurities often strongly disturb or influence 
phase transitions.  Examples of impurities  
  are  filler particles  
 in phase-separating polymer blends \cite{Amis}, 
  microemulsions in nematic 
 liquid crystals \cite{Yamamoto-Tanaka}, 
 and crosslink irregularities in polymer gels 
 \cite{Toyo,Boue,Lube,Onukibook}.
For  gels, some authors developed 
random crosslink models \cite{Lube}.    
Moreover, in gels with liquid crystal solvents \cite{PG,Tere},  
the isotropic-nematic phase transition is  
analogous to the orientation phase transition 
in solids,  where the coupling between 
 the molecular orientation and 
the  elasticity  leads to  
singular elastic behavior. 
  In such liquid crystal gels, 
   nematic polydomains are produced by random crosslinkage 
 and   polydomain-monodomain 
transitions are induced by  
applied stress  or 
electric field \cite{Kai}, as numerically studied 
by Uchida  \cite{Uchida} using  
 quenched random stress. 
The polydomains obviously correspond  to the mesoscopic 
orientation heterogeneities in solids. 
We also mention experiments of 
crystal formation and glass transition 
using elongated   colloidal particles  
in three dimensions \cite{colloid3D} 
and in two dimension \cite{China}.

The mesoscopic heterogeneities 
  produced by impurities 
are  widely recognized   
in various  solid and soft materials. 
We may mention  two previous 
 approaches.  One  is based on a random field 
 coupled to  the order parameter   
\cite{sound, Lube,Uchida,Michel,random-relax,Lookman}.
In particular, Vasseur and  Lookman \cite{Lookman} 
introduced a spin glass theory 
supplemented with  the  elastic 
interaction (the   long-range  interaction among the order 
parameter $\psi$ mediated by the elastic deformations) 
\cite{Onukibook}. 
The other is  a phase-field (Ginzburg-Landau) theory 
with a random  critical temperature  and 
the  elastic interaction  \cite{Kartha,Saxena,SaxenaReview}, 
where the  quadratic term ($\propto \psi^2$)  in 
the  free energy  has a random coefficient.  
In these theories, the impurities are 
  governed by an artificial 
random distribution  without spatial correlations.  
Therefore, they lack     
 microscopic physical  pictures 
of the impurity disordering. 
From our viewpoint,   microscopic approaches 
are particularly needed when each impurity 
strongly  perturbs   the local order parameter. 

To perform  
first-principle  calculations of the mesoscopic 
heterogeneities,  
we  start with   Lennard-Jones systems 
composed of anisotropic host 
particles and impurities  to  create 
orientationally disordered and ordered 
crystal  states.  (i) In such states 
we may examine the degree of heterogeneities  
 by changing   the impurity composition. 
 (ii)  We may describe a tendency of impurity 
 clustering or aggregation \cite{Hama}, which 
   depends   on the cooling rate from liquid. 
 It apparently governs  the degree of vitrification, 
 for example, in water  containing 
 a considerable amount of  salt  
\cite{water}. (iii) We also note that 
the positional disorder 
and the orientation disorder have been discussed 
separately in the literature. 
In this paper,  they  appear simultaneously, though  
 the former is   weaker than the latter.

The organization of this paper is as follows. 
In Sec.II,   we will 
present  the backgrounds of our theory 
and simulation.  
In Sec.III,  we will 
give simulation results 
for  one-component systems  
of elliptic particles forming crystal 
to examine the orientation transition. 
In Sec.IV, we will treat 
 mixtures of elliptic particles and 
larger repulsive impurities, where 
the elliptic particles are aligned  in the planar 
 alignment around the impurities\cite{Prost}.  
In Sec.V, we will examine 
the orientation dynamics of the elliptic particles. 
 In Sec.VI, we will treat small 
attractive  impurities, which 
tend to form aggregates and 
solvate several elliptic particles in the 
homeotropic alignment\cite{Prost}.

\section{Theoretical and simulation backgrounds}

We  propose   a simple microscopic model of   
binary mixtures  in two dimensions, which  exhibits orientation 
 phase transitions  and  glass behavior. 
We do not introduce the dipolar interaction 
supposing nonpolar molecules.

\subsection{Angle-dependent potential}
 
In our model, the first and second 
components are composed  of  elliptic  and spherical 
 particles, respectively. 
Their   numbers are $N_1$ and $N_2$, 
where   $N=N_1+N_2=4096$ in this paper. 
The  composition is defined by 
  \be 
  c=N_2/N, 
\en    
which is either of 0, 0.05, 0.1, 0.15, 0.2, or 0.3 
 in this paper.  Thus the particles of the 
second component constitute impurities. 
The particle   positions are  
written as ${\bi r}_i$ ($i=1, \cdots, N$). 
The orientation vectors of the elliptic particles  
 may be expressed in terms of angles $\theta_i$  as 
\be 
{\bi n}_i=(\cos\theta_i,\sin\theta_i),
\en 
where $i=1, \cdots, N_1$. 

The pair potential $U_{ij}$  between particles 
$i \in \alpha$ and $j\in \beta$ 
($\alpha,\beta=1,2$)  is a  truncated 
modified Lennard-Jones potential. 
That is, for  $r_{ij}> r_c=3\sigma_1$ it is  zero, while  
for  $r_{ij}< r_c=3\sigma_1$ it  reads   
\be  
U_{ij}
=4\ep\bigg[(1+ A_{ij}) 
\frac{\sigma^{12}_{\alpha\beta}}{r_{ij}^{12}}
-(1+ B_{ij}) 
\frac{\sigma_{\alpha\beta}^6}{r_{ij}^6} \bigg] -C_{ij}. 
\en 
Here, $
{\bi r}_i -{\bi r}_j =r_{ij} {\hat{\bi r}}_{ij}$  
 with $r_{ij}= |{\bi r}_{ij}|$. In terms of the  
  diameters $\sigma_{1}$ and $\sigma_2$ of the two species, 
we define   
\be 
\sigma_{\alpha\beta}=
(\sigma_\alpha + \sigma_\beta)/2. 
\en  
In Eq(3),  $C_{ij}$ is  
the value of the first term  at $r=r_c$, 
 ensuring the continuity of $U_{ij}$.  
The $\ep$ is 
the characteristic interaction energy. 

The particle anisotropy 
is taken into account by the 
 angle factors $A_{ij}$ and $B_{ij}$, which depend on 
 the relative direction 
$\hrij= r_{ij}^{-1}{\bi r}_{ij}$ 
and the orientations ${\bi n}_i$ and ${\bi n}_j$ of the 
elliptic particles. There can be a variety of their forms 
depending on the nature of the anisotropic interactions. 
Throughout  this paper, we assume the following form,      
\be
A_{ij} = \chi \delta_{\alpha 1} 
({\bi n}_i\cdot\hrij)^2+\chi 
\delta_{\beta 1} ({\bi n}_j\cdot\hrij)^2, 
\en
where $\chi$ is the anisotropy strength of  repulsion.  The  
 $\delta_{\alpha 1}$ ($\delta_{\beta 1}$) is equal to 
1 for $\alpha=1$ $(\beta=1$) and  0 for $\alpha=2$ ($\beta=2$). 
Thus, in the right hand side, 
 the first (second) term is nonvanishing only when 
$i$ ($j$) belongs to the first species.
In Sec.IV, we  treat 
large spherical impurities repelling 
the elliptic particles by setting  $\sigma_2/\sigma_1>1$ and $B_{ij}=0$.
If $\chi>0$ in  this case, 
there appears a tendency of 
parallel alignment of  the elliptic  particles 
at the impurity surfaces \cite{Prost}.  
On the other hand, in Sec.VI, we  
assume  $\sigma_2/\sigma_1<1$   and 
\be
B_{ij} = \zeta \delta_{\alpha 1}\delta_{\beta 2}
({\bi n}_i\cdot\hrij)^2+
\zeta \delta_{\alpha 2}\delta_{\beta 1}
 ({\bi n}_j\cdot\hrij)^2, 
\en 
where $\zeta$ is the  anisotropy strength of attraction. 
In this case, the  attractive interaction 
is  anisotropic  only between 
the elliptic particles   and 
small  spherical impurities and, 
if $\zeta>0$, there appears a tendency of 
homeotropic  alignment \cite{Prost}   
at the impurity surfaces.

The total energy is written as 
${\cal H}= K+U$, where 
$U$ is the potential energy and $K$ is the kinetic 
energy,  
\bea 
U &=&  \sum_{1\le i<j\le N}U_{ij}, \\
K &=&  \sum_{1\le i\le N}\frac{m_\alpha}{2} |{\dot{\bi r}}_i|^2
+  \sum_{1\le i\le N_1} \frac{I_1}{2} |\dot{\theta}_i|^2,
\ena  
where ${\dot{\bi r}}_i= d{\bi r}_i/dt$,    
${\dot{\theta}}_i = d\theta_i/dt$,  
$m_1$ and  $m_2$   are 
 the masses,   and $I_1$ is the  moment of inertia  
 of the first component. 
In this paper, we set $m_1=m_2=m$.   
The Newton equations  of motion are now written as 
\bea
&&{m_\alpha} 
{\ddot{\bi r}}_i =\pp{}{{\dot{\bi r}}_i}K=-\pp{}{{\bi r}_i}U,
\\
&&{I_1} {\ddot{\theta}}_i = 
\pp{}{{\dot\theta}_i}K= -\pp{}{\theta_i}U, 
\ena 
where  $i\in \alpha$, 
${\ddot{\bi r}}_i= {d^2}{\bi r}_i/{dt^2}$, and 
 ${\ddot{\theta}}_i= {d^2}{\theta}_i/{dt^2}$. 
 The second line holds 
for the first component ($i=1, \cdots, {N_1}$).   
However, since we treat 
equilibrium or  at least 
nearly steady states, 
 we  attach  a Nos$\acute{\rm e}$-Hoover thermostat 
\cite{nose} to all the particles 
by adding the thermostat terms 
in Eqs.(9) and (10). Unless confusion may occur, 
space,   time, and temperature  will be  measured 
in  units of $\sigma_1$,   
\be 
\tau_0 = \sigma_1 \sqrt{m_1/\ep}, 
\en 
and $\epsilon/k_B$, respectively, where $k_B$ is the Boltzmann constant. 
Stress (and elastic moduli) 
will be measured 
in units of $\epsilon/\sigma_1^2$.

From Eqs.(3), (5), and (6)  
the elliptic particles   
have   angle-dependent  diameters. 
Let  the particles $i$ and $j$ belong 
to the first species. Then 
minimization of 
$U_{ij}$ in Eq.(3) with respect to $r_{ij}$ 
 gives $r_{ij}= 2^{1/6}(1+A_{ij})^{1/6}\sigma_1$. 
 Thus the  shortest 
diameter $a_s$   is given for the perpendicular 
orientations  
 (${\bi n}_i\cdot {\hat{\bi r}}_{ij} 
= {\bi n}_j \cdot {\hat{\bi r}}_{ij} =0$), while 
 the  longest  diameter 
$a_\ell$  by the parallel orientations  
(${\bi n}_i\cdot {\hat{\bi r}}_{ij} 
= {\bi n}_j \cdot {\hat{\bi r}}_{ij} =\pm 1$) so that   
\be 
a_s= 2^{1/6}\sigma_{1},\quad  a_\ell=
 (1+2\chi )^{1/6}2^{1/6}\sigma_{1}. 
\en 
 The ratio 
of these lengths (the aspect ratio)  is  
given by $a_\ell/a_s= (1+ 2\chi)^{1/6}$. 
For example, $a_\ell/a_s$ is equal to $1.14$ 
for $\chi=0.6$  and  to 1.23  for $\chi=1.2$. We estimate 
the effective molecular area and 
the momentum of inertia of the elliptic particles 
  as 
\be 
 S_1= 
 \pi a_s a_\ell/4, \quad  
I_1= (a_\ell^2+ a_s^2)  m_1  /4.
\en  
 In this paper, we fix  the overall 
 packing fraction as   
\be  
 \phi_{\rm pack}=(N_1S_1+ N_2 S_2)/V=0.95,
\en   
 where $S_2  =\pi 2^{1/3}\sigma_{2}^2/4$ and $V$ is the 
 system volume.   Then  
the system length  is $L=V^{1/2}\cong 70\sigma_1$. 

Our potential is 
analogous to the Gay-Berne potential 
for anisotropic molecules used 
to simulate mesophases  of liquid crystals \cite{Gay} 
and  the Shintani-Tanaka potential 
with five-fold symmetry 
used to study frustrated particle configurations 
at high densities\cite{Shintani}.  
It is worth noting that 
angle-dependent potentials have been used 
for lipids forming  membranes.  
\cite{Leibler,Noguchi}. 

\subsection{Coarse-grained 
orientation order parameter}

For each particle $i$ 
of the first species ($i=1, \cdots,N_1$), 
 we may   introduce the orientation tensor ${\aQ}_i =
 \{ Q_{i\mu\nu}\}$ ($\mu,\nu=x,y$) 
in terms of the orientation vectors ${\bi n}_k$  as 
\bea
\aQ_i&=&
({1+n_{\rm b}^i})^{-1} ({\bi n}_i{\bi n}_i
 + \sum_{j\in {\rm bonded}} {\bi n}_j {\bi  n}_j) 
-\aI/2 \nonumber \\
&=&q_i ({\bi d}_i {\bi  d}_i -\aI/2),
\ena 
where ${\aI}= \{\delta_{\mu\nu}\}$ is the unit tensor and 
${\bi d}_i$ is  the director with $|{\bi d}_i|=1$.  
The summation is 
 over the bonded particles $(|{\bi r}_{ij}| <3\sigma_1$) 
 of the first species  
with $n_{\rm b}^i$ being the number of 
these   particles of order 20. If a  hexagonal lattice 
is formed, it   includes  
 the  second nearest neighbor particles.  
The   angle  of the director is defined by 
\be 
{\bi d}_i=(\cos\varphi_i,\sin\varphi_i),  
\en 
in the range  $0\le \varphi_i<\pi$.  
The amplitude $q_i$ is  given by 
\be 
q_i^2= 2 \sum_{\mu,\nu}  Q_{i\mu\nu}^2.
\en 
We will calculate the  average over  the elliptic particles, 
\be 
\av{q^2}= \sum_{1\le i \le N_1} q_i^2/N_1, 
\en   
which represents the overall degree of orientation order.  
The angle $\varphi_i$  varies  
  more smoothly than $\theta_i$, but 
  they coincide in ordered domains at low $T$. 
The  $q_i^2$ is of order 0.1 in disordered 
states due to the thermal fluctuations,  
but it increases  up to unity within domains 
at low $T$. 
As a merit in visualization,  $q_i^2$ is  small in the interface regions  
at low $T$ (see the right panels of Fig.1).

Since  the tensor $\aQ_i$ is symmetric and traceless, 
its components  are 
written as $Q_{ixx}= - Q_{iyy}= Q_{i2}/2$ 
and $Q_{ixy}=Q_{iyx}= Q_{i3}/2$. In terms of $\varphi_i$ 
we have  
\be
Q_{i2}=q_i \cos(2\varphi_i), \quad 
Q_{i3}=q_i \sin(2\varphi_i).
\en  
These variables change  
with respect to a rotation of 
the reference frame by  an angle $\psi$ as \cite{Onukibook} 
\bea 
Q_{i2}'&=&Q_{i2}\cos (2\psi) +Q_{i3}\sin(2\psi),\nonumber\\
Q_{i3}'&=&Q_{i3}\cos (2\psi) -Q_{i2}\sin(2\psi).
\ena 
We also  introduce the following density variables  as     
\bea 
Q_2({\bi r})
&=& \sum_{i\in 1} 
Q_{i2} \delta ({\bi r}_i-{\bi r}),\\
Q_3({\bi r})
&=&\sum_{i\in 1} 
Q_{i3}
 \delta ({\bi r}_i-{\bi r}).
\ena 
We will calculate the following structure factor, 
\bea 
S_Q(k)&=& \av{|Q_{2{\bi k}}|^2}\nonumber\\
&=&
 \av{|Q_{3{\bi k}}|^2}, 
 \ena   
where $Q_{2{\bi k}}$ and $Q_{3{\bi k}}$ 
are the Fourier components of 
$Q_{2}({\bi r})$ and $Q_{3}({\bi r})$, respectively. 
From Eq.(20) the structure factor 
of $Q_2$ and that of $Q_3$ 
coincide under  the rotational invariance of the system 
(without stretching), 
leading to the second line of Eq.(23).  
If there is no anisotropic overall 
strain,   the  isotropy holds 
  for $k$ much larger than the inverse system length, 
leading to Eq.(23) and  $\av{Q_{2{\bi k}}Q_{3{\bi k}}^*}=0$.  
 

\begin{figure}[htbp]
\begin{center}
\includegraphics[width=230pt,bb=0 0 176 312]{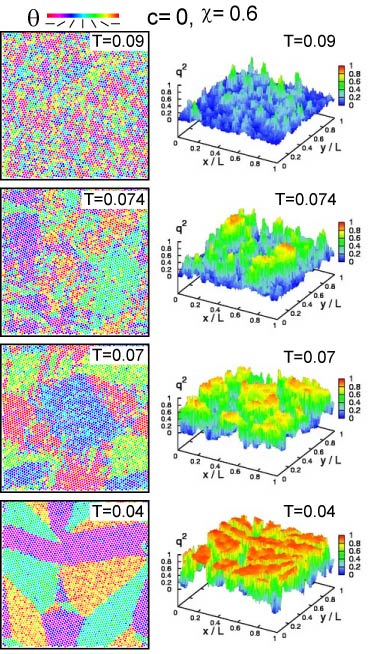}
\caption{Orientation angles $\theta_i$ in the range $0<\theta_i<\pi$ 
(left) and order parameter amplitudes $q^2_i$ (right) 
of all the particles on a lattice 
in the $xy$ plane with 
 $c=0$ and  $\chi=0.6$ for     $T=0.09$, $0.074$, 
$0.07$, and $0.04$ from  above.  As $T$ is  lowered, 
orientation  order 
 develops   gradually with lattice deformations.}
\end{center}
\end{figure}

\begin{figure}[htbp]
\begin{center}
\includegraphics[width=210pt,bb=0 0 278 147]{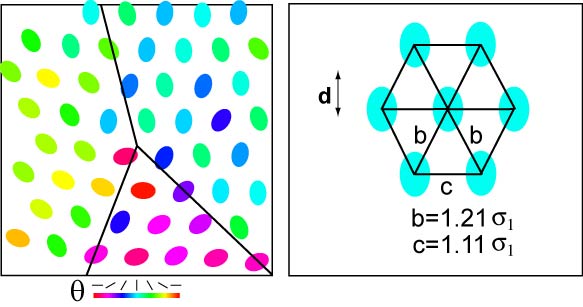}
\caption{Left: Expanded snapshot of  $\theta_i$ 
around  a junction point of three variants in the bottom 
panel of Fig.1 at $T=0.04$.  The angles among  the three lines 
are $65$, $145$, and $150$ degrees, being approximately multiples 
of $\pi/6$. 
Right: Hexagonal lattice structure in an 
ordered  variant 
composed of  isosceles triangles for $\chi=0.6$ and $T=0.04$.}
\end{center}
\end{figure}

\begin{figure}[htbp]
\begin{center}
\includegraphics[width=230pt,bb=0 0 171 102]{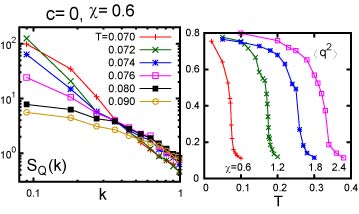}
\caption{Left: 
Structure factor of the orientation fluctuations 
$S_Q(k)$ in Eq.(23) vs $k$ with  $c=0$ and $\chi=0.6$ 
for  $T=0.07,0.072, 0.074,0.076,0.08$, and $0.09$, which  
grows algebraically for small $k <0.3$ in the BKT phase.   Right: 
Average amplitude   $\av{q^2}$ in Eq.(18) vs $T$ 
for $\chi= 0.6$, $1.2$, 
$1.8$, and $2.4$, which increases gradually but 
steeply at low $T$. }
\end{center}
\end{figure}

\section{Orientation 
phase transition in one-component systems}

In this section, we  treat  pure (one-component) 
systems of the elliptic particles ($c=0$).  
We assume not large values of 
 $\chi (\le 2.4)$ such that     
the crystallization first occurs at $T = T_m \sim 1$ 
with random   molecular orientations. 
Far below $T_m$,  
we  study  an   
orientation  phase transition 
on a hexagonal lattice 
and singular mechanical behavior 
specific to  multi-variant states. 
 A number of authors 
  \cite{Frenkel1,Rolf,Kantor}    numerically 
 examined the  phase behavior   of one-component hard rod 
 systems in three dimensions in the plane of 
 the  aspect ratio and the density. 
If the particles are rather close to spheres, 
they found orientationally disordered and 
ordered  crystal phases. Solids in the orientationally 
disordered phase have been called ``plastic solids'' 
\cite{Sherwood,Frenkel1,Rolf,Kantor}.
To understand  singular mechanical 
properties of TiNi 
around its  martensitic phase transition, 
Ding {\it et al.} \cite{Suzuki} performed 
molecular dynamics simulation on   mixtures of 
two species of spherical particles.

\subsection{Variant formation 
and Berezinskii-Kosterlitz-Thouless  phase }

 In  Figs.1-3, 
 we show our simulation results 
 at fixed volume 
  under the periodic  boundary condition. 
  Assuming  a  Nos$\acute{\rm e}$-Hoover thermostat 
\cite{nose}, we started  with 
a liquid  at  $T=2$, 
 quenched the system  to $T=0.35$   
below the melting,     
and annealed it for $9000\tau_0$.  We then  lowered $T$ 
to a final  low temperature. 
Here, even if the cooling rate   was varied 
after  the crystal  formation 
(in the range $T<0.35$), 
 essentially the same results followed. That is,  
there was no history-dependent  behavior.

 In  Fig.1, we show  
the orientation angles $\theta_i$ 
of all the particles in the range $0<\theta_i<\pi$ (left) 
and the order parameter 
amplitudes $q^2_i$ in Eq.(17) (right)  
at $T=0.09, 0.074$, $0.07$, and $0.04$.  
With  lowering  $T$,  
three equivalent  variants emerge 
 due to the underlying  hexagonal lattice. 
Their areal fractions are all nearly equal to $1/3$.  
For $T=0.074$ the time scale of the patterns 
is of order $10^4$, while 
for $T=0.07$ and $0.04$ the patterns are frozen 
even on time scale of $10^5$.   
At  low $T$, 
the junction angles,  at which two or more 
domain boundaries intersect,
are  multiples of $\pi/6$. 
As  illustrated in Fig.2, 
this geometrical constraint  arises from 
the orientation-lattice coupling. It serves to 
  pin  the domain growth   at 
a characteristic size even 
without impurities \cite{Onukibook}. 
  Similar  pinned  domain patterns have been  observed 
on hexagonal planes \cite{Kitano} 
and   were reproduced by 
 phase-field simulation \cite{Chen}.
 We may define the surface tension $\gamma$ 
on the  interfaces   far from 
the junction regions. 
In  our model,  $\gamma\sim 0.1 \epsilon/\sigma_1^2$ 
for $\chi=0.6$ and 
 $\gamma \sim 0.2 \epsilon/\sigma_1^2$ 
for   $\chi=1.2$ at low $T$.

In Fig.3, the structure factor 
$S_Q(k)$ in Eq.(23) vs $k$ and the average $\av{q^2}$ in Eq.(18) 
vs $T$ are displayed for $\chi=0.6, 1.2,1.8$, and 2.4. 
Here, the orientation  order develops continuously  
 in a narrow temperature range,   
\be 
T_2(\chi)< T<T_1(\chi),
\en 
where 
$T_2\sim 0.070$ and $T_1\sim 0.076$ for $\chi=0.6$. 
In our simulation, 
 $T_1$ and $T_2$ increase with increasing $\chi$. 
 They are determined as crossover temperatures. 
In this temperature window,  
   a Berezinskii-Kosterlitz-Thouless  (BKT) phase 
\cite{Jose,Nelson} is realized   
 between the low-temperature  ordered    phase 
 for $T<T_2$ and the high-temperature  disordered  phase for $T>T_1$, 
where the orientation fluctuations 
are   strongly  enhanced at long wavelengths.    
In our model,   each elliptic particle on a lattice point 
behaves as  a rotator in the XY spin  model 
under a symmetry-breaking 
free energy $\Delta F= -\sum_i h_p \cos( p\theta_i)$ 
with $p=6$, 
which arises from the underlying crystal structure \cite{Jose}. 
In accord with the theory \cite{Jose,Nelson}, 
 the structure factor 
 $S_Q(k)$ in Eq.(23) grows algebraically  as 
\be 
 S_Q(k) \sim 
 k^{-2+\eta} \quad (k\ls 0.5)
\en  
in the temperature range (24) with   $\eta$   
 depending  on $T$ ($\eta \cong 0.05$ at $T=0.074$). 
 As regards dynamics, the orientation fluctuations  migrate in space on 
 rather rapid  time scales  slightly below  $T_1$, 
but are frozen  for  $T\ls T_2$ 
(see Fig.13 below). Considerably below $T_2$, 
the three variants become distinct with sharp 
interfaces. Previously, for  two-dimensional hard rods, 
  Bates and Frenkel \cite{Frenkel2}  
found a Kosterlitz-Thouless   phase 
transition   between the  nematic phase 
 and  the  isotropic phase  
 for large aspect ratios and for low densities.

As illustrated in the right panel of Fig.2, 
 the orientation order  induces 
 lattice deformations. 
In ordered states  at low $T$, each  variant  
is composed of  isosceles triangles elongated along 
its orientated direction 
parallel to one of the crystal axes. 
At low $T$, 
their  side lengths $b$ and $c$ are  
$(b,c)= (1.21, 1.11)$ for $\chi=0.6$ 
and $(1.27,  1,11)$ for $\chi=1.2$ 
under the periodic boundary condition, 
while we have $(b,c)=(1.28,1.12)$ at zero stress.
Thus this orientation transition is also 
a structural or martensitic 
 one with spontaneous lattice deformations.

\begin{figure}[t]
\begin{center}
\includegraphics[width=250pt,bb=0 0 167 175]{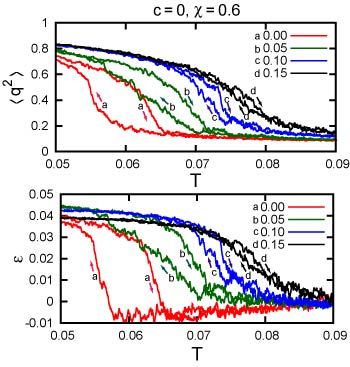}
\caption{Order parameter amplitude $\av{q^2}$ (top) 
and strain $\varepsilon$ (bottom) 
under fixed applied  stress 
$\sigma_{a}=+0$, 0.05, 0.10, and 0.15  
in units of $\varepsilon/\sigma_1^2$, 
where  $c=0$ and $\chi=0.6$. 
The temperature $T$ was 
first decreased from 0.1  
to 0.02 and  it was then increased back to 0.1, where  
 $dT/dt = \mp 4\times 10^{-6}$. 
Hysteretic behavior appears 
between  the  cooling  and  
heating  paths. 
}
\end{center}
\end{figure}

\begin{figure}[t]
\begin{center}
\includegraphics[width=250pt,bb=0 0 307 151]{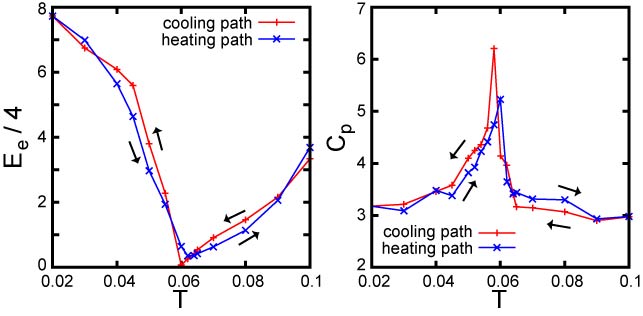}
\caption{Young's modulus 
$E_e$ in Eq.(28) divided by 4  (left) 
and the isobaric specific heat $C_p$ in Eq.(32) (right) 
for $c=0$ on the cooling and heating paths  in Fig.4 
in the nearly stress-free condition ($\sigma_a=10^{-3}$). Here 
$E_e/4$ is nearly equal to 
the effective shear modulus $\mu_e$ in Eq.(29). 
Softening against  shear 
deformations and large energy fluctuations 
are conspicuous at the orientation transition.}
\end{center}
\end{figure}

\begin{figure}[htbp]
\begin{center}
\includegraphics[width=250pt,bb=0 0 286 142]{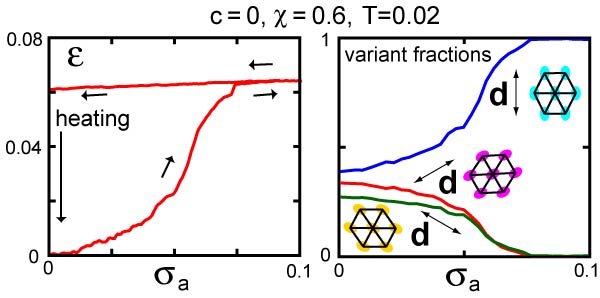}
\caption{Shape memory effect under uniaxial  deformations  
along the $y$ axis without impurities ($c=0$), where 
 $T=0.02$  and $\chi=0.6$. 
Left: Strain $\ve$ vs applied stress ${\sigma_{a}}$ 
in units of $\epsilon/\sigma_1^2$. 
For  ${\sigma_{a}}>0.075$, there remains only 
the variant  elongated along the $y$ axis.
After this cycle, the residual strain 
vanishes  upon heating to  $T=0.1$. 
Right: Fractions of the three variants during  the cycle, 
which are stretched along  the three crystal axes. 
}
\end{center}
\end{figure}

\subsection{Mechanical properties and specific heat 
 for $c=0$}

We have also performed  simulation  
at a fixed stress  \cite{Rahman}, which allows   an 
anisotropic shape change of the system   
at a structural phase transition. 
In  Figs.4-7, we  assumed 
 a Nos$\acute{\rm e}$-Hoover thermostat 
\cite{nose} and a 
Parrinello-Rahman barostat \cite{Rahman} 
under the periodic boundary condition. 
Namely, we controlled  the temperature $T$ and the   
 stress along the $y$ axis written as   
\be 
 \sigma_a= 
\av{\sigma_{yy}}, 
\en  
 where   $\av{\cdots}$ represents the space average. 
The  $y$ axis is taken to be in  
 the perpendicular direction in the figures. 
Hereafter,  $\sigma_a$ will be measured in units of 
$\epsilon/\sigma_1^2$. 
When  $\sigma_a$ was held fixed at a positive value 
for a long time,   a single-variant 
state elongated along the $y$ axis was 
eventually realized  at low $T$. 
This was the case even for 
very small positive $\sigma_a$ ($\sim 10^{-3}$), 
since it serves as a symmetry-breaking 
field. We also carried out  many 
simulation runs    exactly setting   $\sigma_a=0$, 
where a few domains often remained 
in the final state 
depending on the initial 
conditions (not shown in this paper). 
When  $\sigma_a$ is controlled,  the system length 
$L_y$ along the $y$ axis should be  calculated. The 
strain   $\varepsilon$  is defined as  
\be 
\varepsilon= L_y/L_{y0}-1 ,
\en 
where   $L_{y0}$ is a  
reference system length to be specified below.  
We may define Young's  modulus  by     
\be 
E_e=  1/(\p \epsilon/\p {\sigma_{a}})_T.
\en  
even in the nonlinear regime.  
Note that  Young's  modulus is written as 
$E=4K\mu/(K+\mu)$ in the linear regime 
in terms of  the bulk modulus $K$ and  
the shear modulus  $\mu$ in two dimensions. 
We may introduce the effective shear modulus 
$\mu_e$   replacing $E$  by $E_e$ as 
\be 
\mu_e = E_e/(4-E_e/K) \cong E_e/4,
\en 
where we have assumed $E_e\ll K$.

Substantial thermal hysteresis 
during cooling and heating has been observed 
in alloys around 
martensitic phase transitions 
\cite{marten1,marten2,marten3}. Ding {\it et al}. 
also found thermal hysteresis numerically  \cite{Suzuki}.  
In Fig.4, we  show  thermal hysteresis 
in our system for $c=0$. 
That is, fixing  $\sigma_a$, we  
decreased $T$  from 0.1 to 0.02  with 
a very slow cooling rate 
given by $dT/dt= - 4\times 10^{-6}$,  where  
the  variant elongated along the $y$ axis 
became dominant  at low $T$.  
We then increased $T$  back to the initial 
high temperature with $dT/dt=  4\times 10^{-6}$. 
The curves of $\sigma_a=+0$ 
are those with a small symmetry breaking 
stress ($= 10^{-3}$). 
The reference length  
$L_{y0}$ in Eq.(27) is that at  $T=0.09$ 
equal to  $73.0$, 
$73.5$, $75.7$,  and $76.2$ 
for $\sigma_a=+0$, 0.05, $0.1$, 
and  0.15, respectively. 
Hysteretic behavior can be seen in the degree of 
orientation $\av{q^2}$ 
and the strain $\varepsilon$. 
The width of the 
hysteresis  loop  is maximum for $\sigma_a=+0$ 
and shrinks 
to vanish for $\sigma_a>0.15$.  
The   transition at $\sigma_a=+0$ 
between the orientationally disordered 
and ordered states is shifted 
to lower temperatures by 0.01 than in 
the fixed-volume case in Fig.1.  
See Remark (4) in Sec.VII for 
discussions on the stability of quasi-equilibrium 
states in Fig.4.  

When $\av{q^2}$ is small, 
we may use the linear elasticity relations in two dimensions, 
\bea 
&& K [e_1-\alpha (T-T_0)]+ \mu (2\varepsilon-e_1) =\sigma_a, 
\nonumber \\
&& K [e_1-\alpha (T-T_0)]- \mu (2\varepsilon-e_1) =0,  
\ena 
where 
$e_1$ is the dilation strain, 
$\alpha$ is the thermal expansion 
coefficient,   and 
 $T_0$ is a reference temperature . 
The small slope of the curves of $\varepsilon$ 
at $\sigma_a=+0$ in the disordered 
regions  in Fig.4 arise  from the thermal expansion. 
From these relations we obtain 
\bea 
e_1&=& \sigma_a/2K +\alpha (T-T_0),\nonumber\\
\varepsilon&=& (K+\mu)\sigma_a/4K\mu+ \alpha (T-T_0)/2
\ena 
The data in Fig.4 yield 
$K \cong 20$, $\alpha\cong 0.6$, 
and $\mu \cong 2$ with $T_0= 0.09$  
in the disordered phase.  

In the left panel of Fig.5, we show 
Young's modulus $E_e$ in Eq.(28) 
 on the cooling and heating paths of  $\sigma_a=+0$ 
in Fig.4. To calculate it, we superimposed  a small stress 
($=10^{-2})$ to the much smaller   symmetry-breaking  
stress ($=10^{-3})$. Remarkably, 
$E_e$  becomes very small around $T=0.06$ for $c=0$ 
 \cite{Cowley,Onukibook}. 
Similar minimum behavior  of the shear modulus 
has been observed  near the 
orientation and glass transitions 
\cite{ori,sound,sound1}, where the minimum depends 
on the mixture composition.
Previously, using the correlation function 
expression, Murat  and  Kantor \cite{Kantor} calculated 
the elastic constant to find its softening  
toward  the orientation transition 
in two-dimensional ellipsoid systems.  
Nonlinear response  behavior should appear 
 even for very small applied strains 
near the transition. 
Additionally, in the right  panel of Fig.5, we display  
the isobaric specific heat in the nearly stress-free 
condition ($\sigma_a=10^{-3}$) 
along the cooling and heating 
paths expressed as  
\be 
C_p = 
\av{({\cal H}- \av{\cal H})^2}/VT^2,
\en 
 where ${\cal H}=K+U$ is the total energy (see Eqs.(7) and (8)) 
and $\delta{\cal H}={\cal H}-\av{{\cal H}}$ is its 
deviation. 
It  is peaked  at  $T=0.06$ 
indicating  enhancement of the energy fluctuations 
at the transition. 
Such specific  heat anomaly has been measured  
near the orientation transition 
 \cite{Yamamuro}.   We also calculated 
 the constant-volume specific heat $C_V$ 
 using the data in Fig.1 to find a similar peak 
 around $T=0.073$ (not shown in this paper).

 Next,  we illustrate  the shape memory 
effect taking  place 
without dislocations. In Fig.6, we 
increased  ${\sigma_{a}}$ 
from 0 to 0.1 and then decreased ${\sigma_{a}}$  
back to 0   at $T=0.02$, where 
 $d\sigma_a/dt= \pm 7\times 10^{-6}$  
 with  $+$ being  
 on the stretching path and 
 $-$ being  on the return  path. 
 In this slow cycle, the  system remained  in quasi-static 
states.  In the definition of $\varepsilon$  in Eq.(27), 
$L_{y0}$ is the initial  system length $(\cong 72)$. 
 At  $t=0$,   the fractions of the 
three variants  were nearly close to $1/3$  and one 
 variant  was  elongated along the $y$ axis. 
 In the  very early stage 
  $\ve<2\times 10^{-3}$, 
the system deformed elastically with $\mu_e\cong \mu \sim 2$. 
However, in the next stage   $2\times 10^{-3}<\ve< 0.075$, 
 the fraction of the favored variant increased 
 up to unity with $\mu_e \sim 0.1-0.8$.
  This inter-variant 
transformation occurred     
 without   dislocation formation.   
 On the return path,  the solid was composed of 
 the favored variant only with large $\mu_e\sim 7$.     
As  $\sigma_a\to 0$, 
   there remained a  remnant strain about 0.06. However, 
     upon  heating to $T=0.1$ above the  transition,  
it  disappeared  and the solid again assumed  a square shape. 
We note that plastic deformations should occur   
at high strains. 
In the present simulation, 
dislocations  were indeed proliferated 
for ${\sigma_{a}}>0.4$ (or $\epsilon>0.08$)  at $T=0.02$. 


\section{Glass formation  
with large repulsive  impurities }

In Figs.7-11, we treat 
mixtures of  elliptic particles and 
large repulsive impurities. 
With increasing the 
impurity composition $c$, 
the orientation disorder is 
 enhanced and  the long wavelength 
orientation fluctuations are  suppressed, 
resulting in  ``orientation-strain glass''. Here, 
 even  for our anisotropic particle   systems,  
we predict  the  nonlinear mechanical behavior  
studied for strain glass \cite{Ren}.   The BKT phase 
disappears with increasing $c$.

\subsection{Orientation-strain glass}

\begin{figure}[t]
\begin{center}
\includegraphics[width=230pt,bb=0 0 178 242]{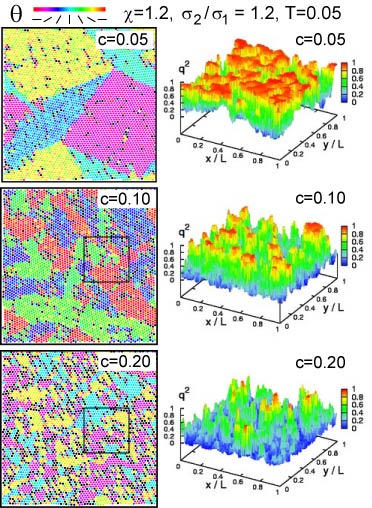}
\caption{Frozen patterns of  angles $\theta_i$ 
in the range $0<\theta_i<\pi$ (left) 
and order parameter amplitudes $q_i^2$ (right) with impurities 
(black points) at  $c=0.05$, $0.1$ and 0.2, where 
 $T=0.05$ and $\chi=1.2$.  
}
\end{center}
\end{figure}

\begin{figure}[h]
\begin{center}
\includegraphics[width=230pt,bb=0 0 188 109]{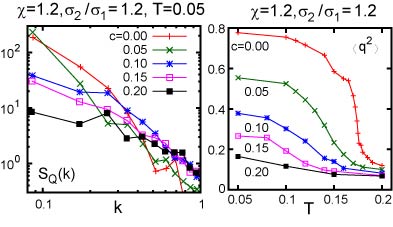}
\caption{$S_Q(k)$ vs $k$ 
with $\chi=1.2$  and $T=0.05$ (left) and  
$\av{q^2}$ vs $T$ with  
$\chi=1.2$ (right). Five curves correspond to 
 $c=0,0.05, 0.1, 0.15$, and 0.2, which 
  are gradually suppressed with increasing $c$.  
}
\end{center}
\end{figure}

\begin{figure}[htbp]
\begin{center}
\includegraphics[width=230pt,bb=0 0 144 93]{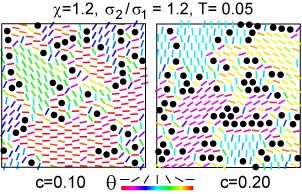}
\caption{Expanded snapshots of  $\theta_i$ 
around large impurities ($\bullet$) for $c=0.1$ and 0.2, 
exhibiting  planar anchoring of 
molecular orientations and clustering.
}
\end{center}
\end{figure}

Figures.7-9 are simulation results 
with a thermostat 
at fixed volume  under the periodic boundary 
condition, where  $T=0.05$  and $\chi=1.2$. 
The temperature was lowered from a high 
temperature as in the previous section. 
The size ratio is fixed at  
 $\sigma_2/\sigma_1=1.2$. 
 For $c\le 0.2$, the system 
still forms a single hexagonal  crystal with  
point defects at the impurity positions.  
 
In Fig.7,  we present  
snapshots of $\theta_i$ and  $q_i^2$    
 for three compositions as in Fig.1.   
In the top panel at  $c=0.05$,  
the impurities induce 
irregular orientation disorder,  
but not much affect the overall  order   
such that large-scale domains are still distinct.  
In the lower panels with  $c=$0.1 and 0.2,  
the orientation  disorder increases and 
the domain sizes become finer. 
For  $c= 0.2$,    
 the system approaches   orientation  
 glass but  with   
  mesoscopically  ordered regions  still remaining. 
In Fig.8,  increasing $c$ 
gives rise to  suppression of  
  $S_Q(k)$ at long wavelengths 
 ($k \ls 0.3$) 
  and $\av{q^2}$ in Eq.(18) at low $T$.

Figure 9 displays  expanded snapshots 
 of the  elliptic particles 
around the impurities. We recognize that 
the   alignments are mostly 
perpendicular  to the surface normals, 
 analogously  to  the  parallel anchoring 
of liquid crystal molecules on the colloid surfaces \cite{Prost}. 
Moreover, we notice an apparent 
 tendency of string-like clustering or 
 aggregation of the impurities. 
They tend to be  
 localized along the interface regions 
between different variants, 
 allowing  formation of  mesoscopically ordered  
 regions of   the elliptic particles 
 even for $c=0.2$. 
 
 To examine the degree of clustering, 
 we may group the  impurities into clusters. 
 Let the two  impurities $i$ and $j$ 
 belong  to  the same cluster 
if their distance is shorter than 
$1.6\sigma_1$. Then we obtain 
the   numbers $N_{\rm cl}(\ell)$ of 
the $\ell$ clusters (those  
consisting of $\ell$ impurities), where  
$\ell=1, 2,\cdots$ and 
 $\sum_\ell \ell  N_{\rm cl}(\ell)=N_2$. 
 The probability that one impurity belongs to 
one of the $\ell$ clusters 
is $P_{\rm cl}(\ell)= \ell  N_{\rm cl}(\ell)/N_2$. 
The average cluster size is defined as 
\be 
{\bar \ell}_{\rm cl}=\sum_{\ell} \ell P_{\rm cl}(\ell)= 
\sum_{\ell} \ell^2   N_{\rm cl}(\ell)/N_2
\en 
In Fig.7, we have 
 ${\bar \ell}_{\rm cl}=1.37$, 2.04, and 4.79 
  for $c=0.05$, 0.1, and  $0.2$, respectively.

\subsection{Mechanical properties in glass}

\begin{figure}[t]
\begin{center}
\includegraphics[width=235pt,bb=0 0 210 351]{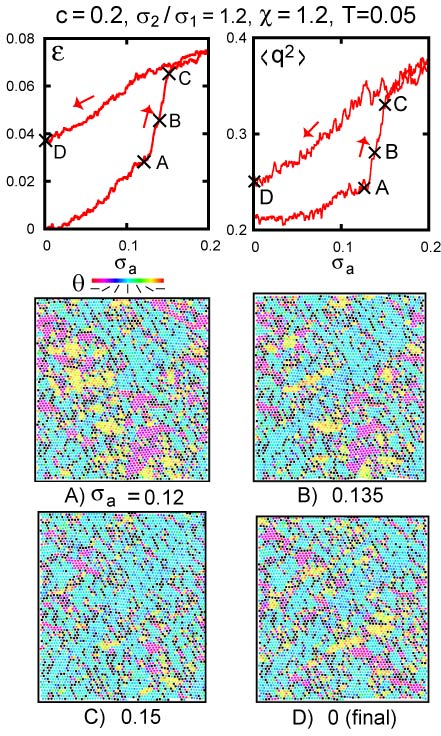}
\caption{Shape memory effect 
under uniaxial deformations 
along the $y$ axis with impurities, where  $T=0.05$, 
 $c=0.2$,  and $\chi=1.2$. 
Top: $\ve$ vs  ${\sigma_{a}}$ (left) 
and $\av{q^2}$ vs ${\sigma_{a}}$ (right). 
For  $0.12<\sigma_a<0.155$, 
 $\varepsilon$ and $\av{q^2}$ 
increase steeply. For ${\sigma_{a}}>0.15$, 
there remains only 
the variant  elongated along the $y$ axis.
After this cycle, the residual strain is 0.04, 
which vanishes  upon heating to  $T=0.1$. 
Bottom: Snapshots of 
 $\theta_i$ for  
$\sigma_a=0.12,0.135$, and 0.15 in the transition region, 
where large-scale orientation fluctuations can be seen 
but there is no dislocation.
 }
\end{center}
\end{figure}

We  also observed a shape-memory effect 
even in orientation    
glass, where small disfavored 
domains  were gradually replaced by  favored 
ones  upon stretching. In this effect, no dislocation 
was formed at low $T$. 

In Fig.10,  at $T=0.05$, 
we increased $\sigma_a$ slowly at 
$d\sigma_a/dt=4\times 10^{-6}$ 
from 0 up to 0.2, where the 
variant elongated along the $y$ axis becomes 
increasingly dominant. 
We then decreased $\sigma_a$   back to 
0 at $d\sigma_a/dt=-4\times 10^{-6}$. 
Between these two paths, significant 
differences can be seen in 
the degree of orientation $\av{q^2}$ 
and the strain $\varepsilon$. 
The $\varepsilon$  
is given by Eq.(27) with 
$L_{y0}$ being  the initial system length 
 at  $T=0.05$ and $\sigma_a=+0$. 
On the stretching path,  there appear four stress 
ranges:  $\mu_e\sim 3$  
for $0<\sigma_a<0.05$,  $\mu_e\sim 0.8$ for 
 $0,05 <\sigma_a<0.12$,   
$\mu_e\sim 0.2$ for $0.12<\sigma_a<0.15$. 
Remarkably, the response is elastic in the first range and is 
very large with   
$\varepsilon$ increasing  steeply from 
$0.028$ to $0.064$ in the third range.  
For $\sigma_a>0.15$ and 
on the return path, $\mu_e$ is of order unity 
and  we can see  considerable variations    
 in $\varepsilon$ and $\av{q^2}$, where 
the fractions of the disfavored 
variants significantly change 
around the impurities.  In contrast, 
  in the one-component case  in Fig.6, 
 we have found  no such changes 
 once a single-variant state is realized.

In the bottom panels of Fig.10, we display 
snapshots of $\theta_i$ at four points A, B, C, and D  
where  $\sigma_a=0.12, 0.135$, 1.5, and 0, 
respectively. See the bottom left panel of Fig.7 
for the snapshot at the initial time 
 in the same run.  
Between  A and B, 
the orientation and the strain  
increase  abruptly. 
In this transition region, we notice  emergence of   
large-scale orientation fluctuations  
taking   stripe shapes and 
 making angles of $\pm \pi/4$ with respect to 
the $x$ axis.  
In stress and thermal cycles in glass, 
  the impurities pin 
 the  orientation fluctuations in  quasi-stationary 
states under very slow time variations of $\sigma_a$ and $T$, 
yielding the history-dependence 
of the physical  quantities.

\subsection{Positional   disorder 
for $\sigma_2/\sigma_1=1.4$ }

 So far, the crystal structure has been     
little affected by the orientation 
fluctuations    at $\sigma_2/\sigma_1=1.2$ 
for $c=0.1$ and 0.2.  
However, if we adopt a  larger size ratio 
and/or a larger composition,   
the  positional  (structural)  
disorder is  increasingly    enhanced, 
resulting in 
usual positional 
 polycrystal  or  glass. 
In our case,   the orientation   disorder 
is  more enhanced than the 
positional disorder. This is in sharp contrast to 
liquid crystal systems where the nematic order precedes 
the crystallization.

 In Fig.11,  
we set $\sigma_2/\sigma_1=1.4$  and $\chi=1.2$ 
to obtain     polycrystal 
  for   $c=0.1$ and 0.2.   In the left,  the orientation angles 
$\theta_j$ are displayed, where  
there still remains noticeable mesoscopic 
orientation  order.  In the left, 
 sixfold bond-orientation (crystal) 
  angles $\alpha_j$ 
are displayed, where 
we introduce  $\alpha_j$ for each elliptic particle $j$ 
in the range $0\le \alpha_j <\pi/3$ by \cite{Nelson,Hama}  
\be
\sum_{k\in\textrm{\scriptsize{bonded}}} \exp [6i\theta_{jk}]
=Z_j \exp[{6i\alpha_j}], 
\en 
 Here,   $\theta_{jk}$ 
is the angle of the relative position vector 
${\bi r}_{jk}= 
{\bi r}_k-{\bi r}_j$ with respect to the $x$ axis, 
 the particle $k$ is within the range 
 $|{\bi r}_{jk}| <1.7\sigma_1$, 
 and   $Z_j$ and $6\alpha_j$ are 
 the absolute value and the phase angle 
 of the left hand side, respectively. 
For $c=0.1$, one large grain is 
embedded in a crystal containing  
many point defects, where 
  angle differences  are 
of order $10-15$ degrees. 
On the other  hand, for $c=0.2$,  many grains appear  
with much larger  angle differences. 

\begin{figure}[t]
\begin{center}
\includegraphics[width=220pt,bb=0 0 160 197]{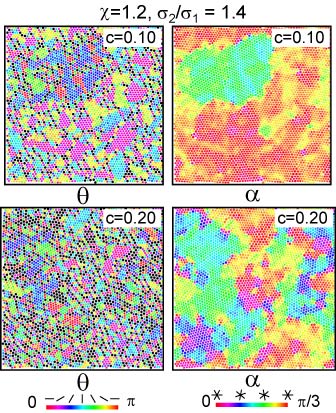}
\caption{
Orientation angle $\theta_i$ in the range $0<\theta_i<\pi$ (left) 
and  sixfold bond orientation 
angle $\alpha_i$ in Eq.(34) in the range $0<\theta_i<\pi/3$ 
(right) in polycrystal states  
for  $c=0.1$ (top)  and 0.2 (bottom),  
where $\chi=1.2$   and $T=0.05$. 
The size ratio is increased to 
$\sigma_2/\sigma_1=1.4$. 
}
\end{center}
\end{figure}

\section{Rotational dynamics}

\subsection{Angle relaxation functions}
 
We now discuss the rotation dynamics of 
the elliptic particles\cite{Chong,new}. 
In two dimensions, we  consider the 
time-dependent angle-distribution function defined by 
\be 
G(t,\varphi)
=\frac{1}{N_1}\sum_{1\le j\le N_1}
 \av{\delta( \theta_j(t+t_0)-\theta_j(t_0)-\varphi)} ,  
\en 
where 
the average $\av{\cdots}$ is taken over 
the initial time $t_0$ and over several runs. 
Here,  $G(t, \varphi)$ tends to 
$\delta(\varphi)$ as $t \to 0$ 
and is  broadened for $t>0$. 
In particular, we treat  the  first two 
moments $G_1(t)$ and $G_2(t)$ written as   
\bea 
G_1(t)&=&    \int_0^{2\pi}{d\varphi}   
G(t,\varphi) \cos(\varphi), 
\\ 
G_2(t)&=&   \int_0^{2\pi} d\varphi  
G(t,\varphi) \cos(2\varphi).  
\ena 
Since these two functions are unity as $t\to 0$, we 
introduce two relaxation times, 
$\tau_1$ and $\tau_2$, by 
\be 
G_1(\tau_1) = e^{-1}, \quad G_2(\tau_2) = e^{-1}.
\en 
These times grow as $T$ is lowered. 
We plot $G_1(t)$ and $G_2(t)$  vs $t$ in  Fig.12 and 
$\tau_1$  vs $T$ in Fig.13.

\begin{figure}[t]
\begin{center}
\includegraphics[width=240pt,bb=0 0 270 279]{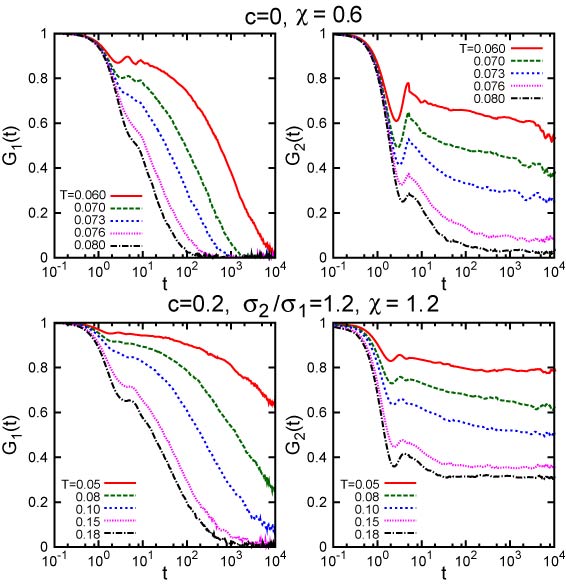}
\caption{Orientation relaxation functions 
$G_1(t)$ (left) and $G_2(t)$ (right) 
in  Eqs.(36) and (37)  for $c=0$ and $\chi=0.6$ (top) and 
for $c=0.2$,  $\sigma_2/\sigma_1=1.2$, 
and $\chi=1.2$ (bottom).  Relaxations  
are slowed down as $T$ is lowered. 
The $G_1(t)$ decays due to 
turnover motions, while  $G_2(t)$ 
due to configuration changes. 
For $c>0$, 
$G_2(t)$  tends to a finite constant as $t\to\infty$.  
 }
\end{center}
\end{figure}

\begin{figure}[t]
\begin{center}
\includegraphics[width=210pt,bb=0 0 150 93]{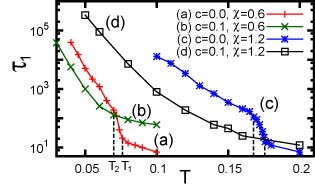}
\caption{Orientation relaxation time $\tau_1$ 
from $G_1(t)$ for (a) $c=0$ and $\chi=0.6$,  (b) 
$c=0.1$ and $\chi=0.6$, 
 (c) $c=0$ and $\chi=1.2$, and  
 (d) $c=0.1$ and $\chi=1.2$. It is 
  the  time scale of successive 
turnover motions.  For (a) and (c), 
 $\tau_1$ grows steeply in the 
BKT  phase ($T_2<T<T_1$).   
For strain glass  (b) and (d), the BKT phase is 
nonexistent and $\tau_1$ 
grows as $T$ is lowered.  
 }
\end{center}
\end{figure}

\subsection{Turnover motions and configuration changes}

As a marked feature, 
the  elliptic particles  sometimes undergo  
the turnover motion    $\theta_j\to  
\theta_j\pm \pi$ or ${\bi n}_j\to  
-{\bi n}_j$ taking  place 
in a  microscopic time $(\sim 1$)\cite{new}.  
In terms of the orientation vector ${\bi n}_j$,  we also have 
\be 
G_1(t) = 
\sum_{1\le j \le N_1} 
\av{{\bi n}_j(t+t_0)\cdot{\bi n}_j(t_0)}/N_1, 
\en 
so that the times between successive 
 turnovers  of an  elliptic particle   
are  of order $\tau_1$ in Eq.(38). 
  On the average over all the elliptic particles,
  the turnover motions  give rise to 
 a peak in   $G(t,\varphi)$ at $\varphi=\pi$. 

In  our simulation,  it is nearly  of the Gaussian form 
for $t\ll\tau_1$  in the range $|\varphi -\pi|\ls 1$ as  
\be 
G(t, \varphi) 
\cong\frac{ A(t)}{\sqrt{2\pi}\sigma} 
\exp\bigg[-\frac{(\varphi-\pi)^2}{2\sigma^2}\bigg],   
\en  
 where the variance $\sigma$ is 
 a constant  about $ 0.45$ in the present case.  
The integral  of this Gaussian peak 
is equal to the coefficient  $A(t)$, so 
$A(t)$ has the meaning of  the turnover probability  
 per elliptic particle 
 in the time interval $[0,t]$.
In terms of $\tau_1$,  
we find the linear growth,
\be  
A(t) \cong C_1  t/\tau_1 , 
\en 
in the early time range  $t\ll \tau_1$. 
In our system $C_1 \cong 0.5$. 
On the other hand, $G_2(t)$ is unchanged 
by the instantaneous turnover motions, 
so it   relaxes    
 due to the orientational 
 configuration changes involving the 
  surrounding particles. We found  
 the inequality  $\tau_2>\tau_1$ 
 at any $T$ and $c$ in our simulation.

In Fig.12, both $G_1(t)$ and  $G_2(t)$ 
relax considerably in the early time region 
$t \ls 2$ due to the thermal rapid motions 
of the orientations without configuration changes. 
For $t\gs 2$ the fitting $G_1(t)  \sim  
\exp[-(t/\tau_1)^\beta]$ 
fairly holds,  
 where  $\beta$ 
 decreases from unity 
 to about 0.5 as $T$ is lowered. 
 Furthermore, for $c>0$,  $G_2(t)$ 
 tends to a nonvanishing positive  
 constant $f_2$ at  large $t$ \cite{Chong}. 
In our case, this plateau appears 
 because the anchoring 
of the elliptic particles 
around the impurities becomes  nearly permanent 
at low $T$. Thus we found that the plateau height 
$f_2$ increases 
with lowering $T$ and with increasing $c$.

In Fig.13,  
the two curves  for $c=0$  
indicate  that 
$\tau_1$ is short ($\ls 10)$ for $T\gs T_1$, 
 increases steeply  
in the BKT region $T_2 \ls T \ls T_1$, 
and  grows further in the ordered  region 
$T \ls T_2$ in the  thermal activation form,   
\be 
\tau_1 \sim \exp(T_0/T) \quad (T \ls T_2).  
\en 
We have  $T_0 \sim 0.40$ at  
$\chi=0.6$ on curve (a) 
and  $T_0 \sim 1.2$ at  $\chi =1.2$ on curve (c). 
In addition, 
 $\tau_1\sim \tau_2$  
for $T \gs T_1$ but $\tau_2/\tau_1 \gg 1$   
for $T\ls T_2$. 
In fact, for $\chi=0.6$,   $\tau_2/ \tau_1$ 
 is about $ 10^2$ 
 at  $T= 0.07$ 
and is about $10^3$ at  $T= 0.06$.  

For $c>0$, the turnover motions 
still occur with $\tau_1<  \tau_2$. 
However, in Fig.13, the relaxation 
behavior for $c>0$  is very different 
from that for $c=0$. 
In the disordered phase with $T\gs T_1$,    
   $\tau_1$ for $c>0$ 
is longer than  $\tau_1$  for $c=0$ 
due to the impurity pinning. 
For $T\ls T_2$,   on the contrary, 
 $\tau_1$ for $c>0$ 
is shorter  than  $\tau_1$ for $c=0$. That is, 
 the turnover motions are more frequent 
  in orientation glass with $c>0$ 
than in the orientationally ordered phase with $c=0$, 
as ought to be the case.  Above $T_2$, the impurity 
anchoring   gradually becomes transient. 

It is worth noting that 
Chong {\it et al} \cite{Chong} 
studied  the  orientation 
dynamics of a glass-forming binary mixture  of dumbbells 
using  the angle relaxation functions 
$C_\ell (t)= \sum_j 
\av{P_\ell({\bi n}_j(t_0+t)
\cdot{\bi n}_j (t_0)}/N$ 
in three dimensional molecular dynamics simulation, 
where $P_\ell$ is the Legendre polynomial of order 
$\ell$  and ${\bi n}_j$ is the orientation vector 
of particle $j$. The relaxations  of 
$C_1 (t)$ and $C_2 (t)$ for small dumbbell anisotropy 
in their paper closely resemble those of 
$G_1(t)$ and $G_2(t)$ for $c=0.2$ 
in Fig.12.

\begin{figure}[t]
\begin{center}
\includegraphics[width=220pt,bb=0 0 147 261]{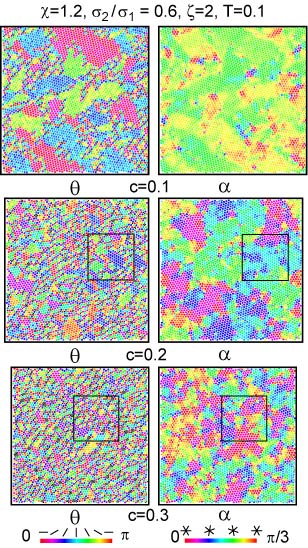}
\caption{Snapshots of orientation angles $\theta_i$ (left) 
and sixfold crystal angles 
$\alpha_i$ with addition of  small attractive impurities 
with $c=0.1$ (top), 0.2 (middle), and 0.3 (bottom), 
where $T=0.1$, $\chi=1.2,$ $\sigma_2/\sigma_1=0.6$, 
and $\zeta=2$.  
The orientation  disorder 
is stronger than the positional disorder. 
The cooling rate from $T=1$ to 0.1 is 
$dT/dt= -1.8 \times 10^{-5}$.}
\end{center}
\end{figure}

\section{
Glass formation  
with small  attractive   impurities }

In this section, we further  treat  
another intriguing case of 
small  attractive  impurities  
with $\sigma_2/\sigma_1=0.6$ 
 in Eq.(3)  and with $\zeta=2$ in Eq.(6).  
Such small impurities tend to be expelled from  the 
ordered domains of the  host particles. 
We shall see that  they   form  clusters.

\subsection{Orientational disorder and positional 
disorder}

Though not shown in this paper,  
we performed simulation runs 
for small repulsive impurities with 
 $\sigma_2/\sigma_1=0.6$ and  $\zeta=0$, where 
 most of the impurity aggregates 
are   stringlike  and     the 
anchoring of the elliptic particles 
is  planar.  
 However, if the anisotropy strength of 
attraction $\zeta$ is increased  at fixed $\chi$,  
the aggregates becomes increasingly compact. 
For $\zeta\gs 1$, the aggregates  can ``solvate''  several 
elliptic particles in the  homeotropic alignment\cite{Prost}. 
With further increasing $\zeta$, 
even a single  impurity  creates  
  a solvation shell composed of 
several elliptic particles like a small metallic ion in water.  

\begin{figure}[t]
\begin{center}
\includegraphics[width=220pt,bb=0 0 139 178]{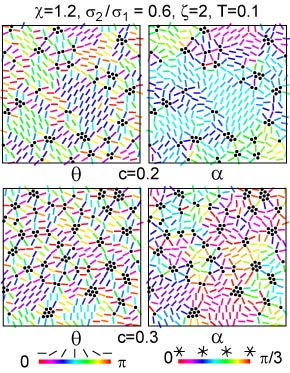}
\caption{Expanded snapshots of 
 elliptic particles 
and  small attractive impurities 
(black points) in the boxes in Fig.14, where 
homeotropic anchoring and impurity clustering are marked.  
Colors of the elliptic particles 
represent the orientation angles 
$\theta_j$ (left) and the sixfold crystal angles 
$\alpha_j$ (right) from the same data.   
The  domains (left) are finer than the grains (right). 
The cooling rate from $T=1$ to 0.1 is 
$dT/dt= -1.8 \times 10^{-5}$.}
\end{center}
\end{figure}

In Fig.14, we  show snapshots of $\theta_j$ 
and $\alpha_j$ of all the particles, 
 where $T=0.1$, $\chi=1.2,$ $\sigma_2/\sigma_1=0.6$, 
 and $\zeta=2$. 
Here, we set $dT/dt= -1.8 \times 10^{-5}$.  
For $c=0.1$,  the system is still in a single 
crystal state, but  
 the  orientational domain structure 
 induces  large-scale  elastic  deformations, 
 leading to close resemblance of 
  the patterns of  $\theta_j$ 
and $\alpha_j$.  
 For $c=0.2$, the orientational domains are much finer 
 and a polycrystal state  is realized with 
 larger  grains ($\gs 10$).  
   For $c= 0.3$, the orientation order 
  is much more suppressed and a positional glass state 
  is realized with mesoscopic 
  heterogeneities  still remaining.

In Fig.15, we display  expanded snapshots 
of  $\theta_j$ (left) and $\alpha_j$ in Eq.(34) (right) 
in the box regions in Fig.14. The  
 alignments of the  elliptic particles 
around the impurities are mostly 
parallel to the surface normals. 
This is analogous to  the  homeotropic  anchoring 
of liquid crystal molecules on the colloid surfaces \cite{Prost}. 
We notice a  tendency of  clustering or  
 aggregation of the impurities. 
Comparing the left and right panels, 
we recognize that the interfaces are finer 
than the grain boundaries. 
That is, the interfaces can be seen 
both on  the grain boundaries 
and within the grains. 
The impurities tend to be  
 localized on  the interface regions 
between different variants.

\begin{figure}[t]
\begin{center}
\includegraphics[width=220pt,bb=0 0 149 113]{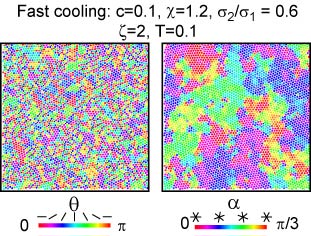}
\caption{Snapshots of 
 orientation angles $\theta_i$ (left) 
and sixfold crystal angles 
$\alpha_i$ (right) with addition of  
small attractive impurities 
with $c=0.1$.   
The   parameter values are 
common to those in the top panel of Fig.14, but the 
 cooling rate from $T=1$ to 0.1 is 
$dT/dt= -9\times 10^{-3}$.   Here,  
the degree of clustering is weaker, 
resulting in 
 a  polycrystal state. 
}
\end{center}
\end{figure}

\subsection{Cooling-rate dependent clustering 
of impurities}

\begin{figure}[htbp]
\begin{center}
\includegraphics[width=200pt,bb=0 0 140 122]{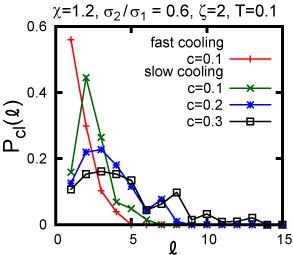}
\caption{ Probability 
$P_{\rm cl}(\ell)= \ell   N_{\rm cl}(\ell)/N_2$ 
of an impurity belonging to $\ell$ clusters, 
where $ N_{\rm cl}(\ell)$ is the cluster 
number composed of $\ell$ impurities. Here 
small attractive impurities 
are considered for slow cooling  
in Fig.14 and  fast cooling in Fig.16.  
Distribution is broader for slower cooling. 
}
\end{center}
\end{figure}


The degree of impurity clustering should be decreased 
 with increasing 
the cooling rate $dT/dt$ for long diffusion times of impurities. 
In Fig.16,  $dT/dt$   is  $ -9 \times 10^{-3}$ 
and is  500 times faster  than in Fig.14, 
where the other parameters are common. 
We give snapshots of $\theta_j$ and $\alpha_j$  
at  $c=0.1$, where  the clustering 
can be  more evidently seen  than  for  $c=0.2$ and 0.3.  
While a single crystal has been  realized  
in the top panel of Fig.14, 
a polycrystal state is realized 
with large angle differences in Fig.16.

Let the two small  impurities $i$ and $j$ 
 belong  to  the same cluster 
if their distance is shorter than 
$1.2\sigma_1$. Then we obtain the  
number $ N_{\rm cl}(\ell) $ of clusters  
composed of $\ell$ impurities.
In Fig.17, we show the cluster size distribution 
$P_{\rm cl}(\ell)= \ell   N_{\rm cl}(\ell)/N_2$ ($\ell=1,2,\cdots$) 
for the examples in Figs.14 and 16. 
The average cluster size  $\bar{\ell}_{\rm cl}$ 
in Eq.(33)  increases with $c$ as $2.45$ for $c=0.1$, 
3.43 for $c=0.2$, and  4.61 for $c=0.3$ under  the slow  
cooling in Fig.14, while 
$\bar{\ell}_{\rm cl}=1.62$  for $c=0.1$ 
under the fast cooling in Fig.16.


It is known \cite{water} that water becomes 
glass at low $T$ with addition of 
a considerable amount of LiCl, where small 
hydrophilic Li$^+$ and Cl$^-$ ions 
solvate several water  
molecules via the strong ion-dipole 
interaction.  The resultant orientation  anchoring 
 of water molecules should even prevent   
formation of the crystal order at high salt 
concentrations, resulting in the observed positional  glass. 
It is natural that the cooling rate  
influences  the degrees of 
ion clustering and  vitrification.

\section{Summary and remarks}

We have presented an   angle-dependent 
Lennard-Jones potential for elliptic 
particles and impurities, which depends on 
the orientation angles of the interacting  particles. 
Using this potential, 
we have performed simulation of 4096  particles 
on very long time scales ($\sim 10^5\tau_0$) 
in two dimensions. 
Our main results are as follows.\\
(i) In Sec.II, we have  presented 
our model potential, where the 
anisotropy strengths are characterized  by 
$\chi$  for the repulsive part 
in Eq.(5) and $\zeta$ for the attractive 
part in Eq.(6). The aspect ratio of the elliptic particles 
 is  given by $a_\ell/a_s= 
 (1+2\chi)^{1/6}$.   In this paper, 
  $\chi$ is of order unity, so we have assumed weak  
  particle anisotropy to find  crystallization 
  at a high temperature above the orientation transition.  
  \\ 
 (ii)In Sec.III, we have presented simulation results for 
one-component systems of elliptic particles 
 by changing the 
temperature $T$ to produce Figs.1-6. 
The domain patterns in Fig.1 at low $T$ are those 
observed on hexagonal planes. 
 In our case, the  Berezinskii-Kosterlitz-Thouless 
  phase\cite{Jose,Nelson} is realized 
in a temperature window, where 
 the orientation fluctuations are 
 much enhanced at long wavelengths as 
 indicated by the structure factor $S_Q(k)$ in Fig.3. 
 We have shown  thermal hysteresis in Fig.4, 
  singular behaviors of 
 the shear modulus and 
 the specific heat  in Fig.5, and a shape-memory 
 effect in Fig.6.  
 \\  
(iii) In Sec.IV, we have examined 
orientation-strain   glass of 
 elliptic particles and large 
repulsive impurities with the size ratio 
$\sigma_2/\sigma_1=1.2$ in Figs.7-10.  The orientations 
of the elliptic particles are 
pinned at the impurity surfaces in the planar 
alignment in Fig.9. The shape-memory effect in strain glass 
is marked in Fig.10.   
Positional disorder  also emerges  
for $\sigma_2/\sigma_1=1.4$ 
 in Fig.11. 
\\
(iv) In Sec.V, we have studied the rotational dynamics 
of the elliptic particles. 
In Fig.12,  $G_1(t)$ decays 
due to the turnover motions of 
 the elliptic particles, while 
 $G_2(t)$ decays due to 
 the configuration changes.  In Fig.13, 
 the turnover relaxation time $\tau_1$ grows at low $T$ 
 and behaves differently with and without impurities. 
\\ 
(v) In Sec.VI, we have examined 
 the effect  of  small attractive  impurities 
 on the orientation disorder 
 and the positional disorder in Fig.14. 
The impurity effect is stronger 
on the former than on the the latter.  
The elliptic particles are 
homeotropically anchored 
at the impurity surfaces in Fig.15. 
The clustering of impurities 
is  suppressed for rapid cooling 
as in Figs.16 and 17. 
\\

We further make critical remarks as follows:\\
(1) In our simulation, we used 
 a Nos$\acute{\rm e}$-Hoover thermostat (NHT) 
\cite{nose} in all the figures and  
a NHT and a Parrinello-Rahman barostat \cite{Rahman} 
in Figs.4-6, and 10. In future work, 
we should examine the coupled 
dynamics of the translational  and  
orientational degrees of freedom \cite{Bell} without 
thermostats and barostats in the system interior.\\  
(2) There has been no systematic 
measurement  of the mechanical properties 
of orientationally  ordered, 
  multi-variant crystal 
and orientation glass. Such experimental results  could 
be compared with those from shape-memory alloys 
\cite{marten1,marten2,marten3,RenReview,Ren}. 
Weak elasticity was observed in orientationally 
disordered solids above the transition 
(called ``plastic solids'') in creep experiments \cite{Sherwood}. 
Also, as far as the authors 
are aware, there has been  no experimental information 
of the impurity clustering in any physical 
systems exhibiting mesoscopic heterogeneities.  
\\ 
(3) In this paper, the particle anisotropy is not large, 
which favors formation of crystal order. 
For large anisotropy,  
liquid crystal  phases should appear \cite{Frenkel1,Rolf}, 
where the impurity   effect  is  of great interest. 
As suggested by the experiment \cite{Yamamoto-Tanaka}, 
addition of a considerable amount of  
impurities leads to the orientation order  
only on mesoscopic scales in liquid crystal 
phases.  In such states, we expect large 
response to applied electric field. \\ 
(4) 
In Figs.1 and 4, our system undergoes  
a structural phase transition 
 gradually in a narrow temperature window 
even for the one-component case. 
In our model,  a  gradual phase transition 
still occurs  in the stress-free condition 
  without impurities. 
However,  we also stopped the cooling 
and waited for a long time ($\gg 10^4$) at $T=0.06$ 
on the stress-free cooling path in Fig.4; then,  
we observed a transition to 
the ordered single-variant phase 
(not shown in this paper). 
Thus, in future work, 
 we need to calculate the Gibbs or Helmholtz 
 free energy to decide whether the system is 
 in equilibrium or in a metastable state.\\
(5)    
As well as the orientation fluctuations, 
the displacement fluctuations are 
also  enhanced around the orientation transition, 
as indicated by  
Fig.5 and by the previous experiments \cite{ori,sound,sound1}. 
In addition,  according to  Cowley's   
classification of elastic instabilities  \cite{Cowley},  
  our phase transitions belong to 
type-I instabilities where 
acoustic modes become soft in particular 
wave vector directions.\\ 
(6) The disordering effect induced by impurities  
 prevents  a sharp transition \cite{sound}. 
Thus there is  no sharp phase boundary 
between the high-temperature orientationally disordered 
phase and the low-temperature orientation-strain 
 glass phase. These two phases  change over gradually 
 with varying $T$ as in the cases of 
  positional glass transitions.  
\\ 
(7) 
Ding {\it et al.} \cite{Suzuki} numerically studied  
the superelasticity, which   arises  
from a stress-induced martensitic 
 phase transition \cite{RenReview,Ren}. 
We also realized this phenomenon for  
 anisotropic particles, 
which will be reported elsewhere.\\ 
(8)   We will also  report 
 three-dimensional 
simulation  on mixtures of spheroidal particles and 
spherical ones without and with 
the dipolar interaction. 
 We shall see  finely divided  domains 
 produced by impurities and 
 large responses to applied strain and 
 electric field.

\begin{acknowledgments}
This work was supported by Grant-in-Aid 
for Scientific Research  from the Ministry of Education, 
Culture,  Sports, Science and Technology of Japan. 
The authors would like to thank Takeshi Kawasaki, 
Osamu Yamamuro,   Hajime Tanaka, 
and  Hartmut L$\ddot{\rm o}$wen    for 
informative discussions. The numerical 
calculations were carried out on SR16000 at YITP in Kyoto
University.
\end{acknowledgments}

\end{document}